\documentclass[aps,prd]{revtex4}

\usepackage{amsmath}

\begin{document}

\title{Modified model of top quark condensation}

\author{M.A.~Zubkov\footnote{on leave of absence from ITEP, B.Cheremushkinskaya 25, Moscow, 117259, Russia}}
\affiliation{University of Western Ontario,  London, ON, Canada N6A 5B7}

\begin{abstract}
We develop the modification of the top - quark condensation scenario, in which the Higgs boson is composed of all Standard Model fermions. Within this scenario we suggest the phenomenological model with non - local four - fermion interactions in which at the distances of the order of $\sim 1/100$ GeV the theory is represented in terms of only one Majorana spinor that carries the $U(12)$ index and, in addition, belongs to the spinor representation of $O(4)$. The Standard Model fermions are the components of this spinor. The symmetry $U(12)\otimes O(4)$ is responsible for the one - loop relation between the Higgs boson mass and the top quark mass $M_H^2 = m_t^2/2$. At the distances $\gg 1/100$ GeV the mentioned symmetry is broken, and the interaction term dominates that provides the nonzero mass of the top quark. Our phenomenological model should be considered in zeta or dimensional regularization. In conventional cutoff regularization this is equivalent to the existence of the additional counter - terms that cancel all quadratic divergences. As a result the $1/N$ expansion may be applied.
\end{abstract}



\maketitle



\section{Introduction}

The original idea of top - quark condensation \cite{Miransky,Nambu,Marciano:1989xd,topcolor1,Hill:1991at,Hill:2002ap,Miransky:1994vk} implied that the Higgs boson is composed of the top - quark. In \cite{ZetaHiggs} we suggested the modification of this  scenario, in which the $125$ GeV h - boson \cite{CMSHiggs,ATLASHiggs} is  composed of all known quarks and leptons of the Standard Model (SM) (see also \cite{Terazawa}, where it was suggested, that the Higgs boson is composed of the SM fermions). In our approach the W and Z boson masses as well as the $h$ - boson mass and SM fermion masses are determined by the condensate of the $125$ GeV $h$ - boson according to the Higgs mechanism \cite{Englert,Higgs}. The important difference from the conventional models of top - quark condensation is that the new strong dynamics is rather complicated and is not described by the point - like four fermion interaction \cite{NJL}.   Unlike the conventional models, where the scale of the new dynamics was assumed to be at about $\Lambda \sim 10^{15}$ GeV, in our scenario the scale of the new strong dynamics is supposed to be of the order of several TeV. The conventional top - quark condensation models typically predict the Higgs boson mass not very different from $2 m_t \approx 350$ GeV, and are excluded by the present experimental data \footnote{In those models the prediction of the Higgs boson mass is the subject of the large renormalization group corrections due to the running of coupling constants between the working scale $\Lambda$ and the electroweak scale $\sim 100 $ GeV. But this running is not able to explain the appearance of the small Higgs boson mass around $125$ GeV.}. It is worth mentioning that recently the modification of the top - condensation scenario was suggested \cite{topcondmod}, in which the $125$ GeV Higgs boson appears as the pseudo - Goldstone boson. In this scenario the value of the Higgs boson mass is suppressed naturally, but the inclusion of extra fermions is necessary.

In \cite{ZetaHiggs} it was shown that the  non - trivial form - factors for the interaction between the Higgs boson and the SM fermions are able to provide the composite Higgs boson mass $M_H = m_t/\sqrt{2} \approx 125$ GeV provided that {\it at the distances $\sim 1/100$ GeV} all SM fermions interact with the composite Higgs boson field in an equal way. Following this scenario in the present paper we suggest the particular model, in which
all SM fermions are arranged within one Majorana spinor. This Majorana spinor carries the $U(12)$ index. In addition, it belongs to the $4D$ spinor representation of $O(4)$. The fermions of different $O(4)$ chiralities may be transformed to each other with the emission of the $125$ GeV Higgs boson.
The Higgs boson in this model appears as the real - valued vector from the representation of $O(4) \simeq SU(2)_L\otimes SU(2)_R$. In the SM only the $SU(2)_L$ component and the $U(1)_Y \subset SU(2)_R$ component of $O(4)$ are gauged. Using the $SU(2)_L$ transformations we are able to bring the Higgs field to the simple form with only one non-vanishing real component (the analogue of the Unitary gauge of the Standard Model).

{\it At the distances  much larger, than $1/100$ GeV} the interaction becomes more complicated. We consider the simplified scenario, in which the global symmetry $O(4) \otimes U(12)$  of the interaction between the composite Higgs boson and the fermions is broken, and one of the possible interaction terms dominates that provides the top  - quark mass. The other fermions remain massless on this level of understanding. We assume, that they acquire masses due to the perturbations above the considered pattern.

The low energy effective model with the four - fermion interaction is not renormalizable. It is the effective theory only and its output strongly depends on the regularization scheme. We imply the use of zeta regularization (see, for example,  \cite{zeta,McKeon}) or dimensional regularization (see, for example,  \cite{NJLdim}) to give sense to the expressions for the observables.
The $1/N$ expansion works good enough in the effective theory because of the chosen regularization. This is well - known, that the ultraviolet divergences break the $1/N$ expansion in the NJL models defined in ordinary cutoff regularization (see, for example, \cite{cvetic}). However, in zeta regularization the ultraviolet divergences do not appear at all while in dimensional regularization only the logarithmic divergences appear while the quadratic ones are absent. This is the reason, why the leading order $1/N$ approximation to various quantities does gives the reasonable estimates. The zeta/dimensional - regularized NJL model is equivalent to the NJL model in ordinary cutoff regularization with the additional counter - terms that cancel all quadratic divergences. As a result of this subtraction, in particular, the sign of the four fermion coupling constant is formally changed. For the relation between the values of this coupling constant before and after the subtraction see, for example, \cite{ZetaNeutrino}, Appendix, Sect. 4.2. That's why the attractive four - fermion interaction in the bare lagrangian of zeta/dimensionally regularized model has the unusual sign (that naively looks like that of the repulsive interaction). There, presumably, exists the class of renormalizable theories with attractive interaction between the fermions that are approximated well by such NJL models defined in zeta/dimensional regularization. We imply, that the theory standing behind the SM belongs to this class of theories. Therefore, while considering the four - fermion approximation to this unknown theory in the present paper we rely on zeta/dimensional regularization. Notice, that the existence of such theories, in which the quadratic divergences of the NJL approximation are to be subtracted was mentioned in \cite{VZ2012}, where this subtraction was related to the existence of a certain stability principle similar to that of the condensed matter theories where the divergences in the vacuum energy of the hydrodynamic description are subtracted by the complete theory due to the thermodynamical stability of vacuum (see \cite{VolovikKlinkhamer2008,KlinkhamerVolovik2010,quantum_hydro,hydro_gravity} and recent review \cite{Volovik2013}).

It is worth mentioning, that there were attempts to incorporate the existence of several composite Higgs bosons $M_{H,i}$ related to the top - quark mass by the  Nambu Sum rule $\sum M^2_{H,i} = 4 m_t^2$. Those models admit the appearance of the $125$ GeV $h$ - boson and  predict the existence of its Nambu partners \cite{VZ2012,VZ2013,VZ2013_2,Z2013JHEP}. The Nambu sum rule  gives $M^2_H = 4m_t^2$ in the simplest model of top - quark condensation \cite{Miransky,Nambu,Marciano:1989xd,topcolor1,Hill:1991at,Hill:2002ap,Miransky:1994vk}, where there is only one Higgs boson. In our model there is also the only Higgs boson in the given channel, but the Nambu sum rule is changed: we have $N_{\rm total} M_H^2 = 4N_c m_t^2$, where
$N_{\rm total} = 2\times (N_c + 1) \times N_g=24$,  $N_c = 3$ is the number of colors, $N_g = 3$ is the number of generations.

\section{The model at the distances $\sim 1/100$ GeV. The Standard Model fermions as the components of the only Majorana spinor.}
\label{Sectfermions}
We adopt the notations used in \cite{ZetaHiggs}. For the completeness we describe them here briefly.
Left handed doublets and the right - handed doublets of quarks are denoted by $L_K^{\bf a}$ and $R_K^{\bf a}$, where $\bf a$ is the generation index while $K$ is the color index. The
left handed doublets and the right - handed doublets of leptons are ${\cal L}^{\bf a} $ and ${\cal R}^{\bf a} $ respectively.
It will be useful to identify the lepton of each generation as the fourth component of colored quark. Then $L^{\bf a}_{a,4} = {\cal L}^{\bf a}_a$ and $R^{\bf a}_{a,4} = {\cal R}^{\bf a}_a$. So, later we consider the lepton number as the fourth color in the symmetric expressions. We define the analogue of the Nambu - Gorkov spinor
\begin{equation}
{\bf L}^{{\bf a}A}_{aiU} = \left(\begin{array}{c}L^{{\bf a}A}_{ai}\\ \bar{L}^{{\bf a}B}_{c^{\prime}i}\epsilon_{c^{\prime}a}\epsilon^{BA}\end{array} \right)
, \, {\bf R}^{{\bf a}A}_{aiU} = \left(\begin{array}{c}\bar{R}^{{\bf a}B}_{bi}\epsilon_{ba}\epsilon^{BA}\\ R^{{\bf a}A}_{a,i} \end{array} \right),\nonumber
\end{equation}
where $A$ is the usual spin index, $U$ is the Nambu - Gorkov spin index ($U = 1,2$ and $A = 1,2$),  $i$ is the $SU(4)$ Pati - Salam color index (the lepton number is the fourth color), $\bf a$ is the generation index, and $a,b$ are the $SU(2)_L, SU(2)_R$ indices. Both ${\bf R}^{{\bf a}A}_{aiU}$ and ${\bf L}^{{\bf a}A}_{aiU}$ for the fixed values of $a$, $i$, and $\bf a$ compose the four - component Dirac spinors ${\bf R}^{{\bf a}}_{ai}$ and ${\bf L}^{{\bf a}}_{ai}$.
These spinors for the fixed value of $a$ have $(N_c + 1) \times N_g=12$ components. Both ${\bf R}$ and ${\bf L}$ belong to the fundamental representation of $U((N_c + 1) \times N_g)$, where $N_c = 3$ is the number of colors, $N_g = 3$ is the number of generations. Notice, that ${\bf L}$ and $\bar{\bf L}$ ($\bf R$ and $\bar{\bf R}$) are not independent:
$\bar{\bf L}^{{\bf a}}_{ai} = \epsilon_{ab} \Bigl( {\bf L}^{{\bf a}}_{bi}\Bigr)^T i \gamma^2 \gamma^5\gamma^0, \,
\bar{\bf R}^{{\bf a}}_{ai} = \epsilon_{ab}\Bigl( {\bf R}^{{\bf a}}_{bi}\Bigr)^T i \gamma^2 \gamma^5\gamma^0$.


Next, we  arrange the Dirac spinors ${\bf L}^{\bf a}_{ai}, {\bf R}^{\bf a}_{ai}$ in the $SO(4)$ spinor $\Psi$:
\begin{equation}
\Psi^{\bf a}_i = \left(\begin{array}{c} {\bf L}^{\bf a}_{ai}\\{\bf R}^{\bf a}_{ai}\end{array}\right)\nonumber
\end{equation}
We introduce the Euclidean $SO(4)$ gamma - matrices $\Gamma^a$ (in chiral representation). The action of the SM gauge fields $e^{i \theta}\in U(1)_Y \subset SU(2)_R$, $U^{(L)}\in SU(2)_L$, $U^{(R)}  = \left(\begin{array}{cc} e^{i\theta}& 0 \\0 & e^{-i \theta} \end{array} \right)\in SU(2)_R$  and $V = \left(\begin{array}{cc}Q e^{i\theta/3}& 0 \\0 & e^{-i \theta} \end{array} \right)\in SU(4)_{\rm Pati\, Salam} \subset U(12)$ (where $Q\in SU(3)$) on the given Majorana spinor is:
\begin{eqnarray}
{\Psi}^{\bf a}_{i} &\rightarrow & \Bigl(V_{ij}\frac{1+\Gamma^5\gamma^5}{2} + \bar{V}_{ij}\frac{1-\Gamma^5\gamma^5}{2}\Bigr)  \left( \begin{array}{cc} U^{(L)}_{} & 0 \\ 0 & U^{(R)}_{}\end{array}\right)  {\Psi}^{\bf a}_{j}\nonumber
\end{eqnarray}
Thus $U^{(L)}, U^{(R)}$ realize the representation of $O(4)\simeq SU(2)_L \otimes SU(2)_R$ while $V$ realizes the representation of the subgroup $SU(4)$ of $U(12)$. The action of the element $R \in U(12)$ of the latter group on the spinor $\Psi^{\bf a}_i$ is $\Psi^{\bf a}_i \rightarrow \Bigl(R^{\bf ab}_{ij}\frac{1+\Gamma^5\gamma^5}{2} + \bar{R}^{\bf ab}_{ij}\frac{1-\Gamma^5\gamma^5}{2}\Bigr)  \Psi^{\bf b}_j$.

Again, $\Psi$ and $\bar{\Psi}$ are not independent:
$\bar{\Psi}^{{\bf a}}_{i} = \Bigl( {\Psi}^{{\bf a}}_{i}\Bigr)^T i \gamma^2 \gamma^5\gamma^0 \Gamma^4 \Gamma^2 \Gamma^5$.
The partition function for the SM fermions in the presence of the SM gauge fields has the form:
$Z = \int D {\Psi} e^{iS}$.
The action $S =  S_K + S^{(4)}_{I}$ contains two terms. The first one is the kinetic term
\begin{equation}
S_K = \frac{i}{2}  \int d^4x\Bigl(\bar{\Psi}^{{\bf a}}_{i} \gamma^{\mu} \nabla_{\mu}{\Psi }^{{\bf a}}_{i}  \Bigr)
 \label{I4l_}
\end{equation}
One can check that this term being written in terms of the original SM fermions is reduced to the conventional SM fermion action (without mass term). Here $\nabla_{\mu}$ is the covariant derivative that includes the gauge field of the model.

$S^{(4)}_I$ is the four - fermion interaction term
\begin{equation}
S^{(4)}_I =  \frac{1}{16 M_I^2}\int d^4x\Bigl(\bar{\Psi}^{{\bf a}}_{i}\gamma^5 \Gamma^5 \Gamma^K {\Psi}^{{\bf a}}_{i}  \Bigr)\Bigl(\bar{\Psi}^{{\bf b}}_{j}\gamma^5 \Gamma^5 \Gamma^K {\Psi}^{{\bf b}}_{j}  \Bigr)\label{SI04}
\end{equation}
As it was already mentioned in the Introduction, we assume that this term appears in the effective low energy description of a renormalizable theory. According to our supposition this low energy description should be considered with all quadratic divergences subtracted, i.e. in zeta or dimensional regularization. In this case the $1/N$ expansion is applicable to the effective NJL model, and we are able to estimate the effective action in one loop.
The given four - fermion interaction describes the dynamics at the electroweak scale  but already is not relevant at the energies much smaller, than $100$ GeV.

As usual, the auxiliary scalar field of the Higgs boson may be introduced. In our case it appears in the form
\begin{equation}
{\bf H} = \sum_{K = 1,2,3,4} {\bf h}_K \Gamma^K\nonumber
\end{equation}
where ${\bf h}^K \in {\cal R}$. Thus, in our model the Higgs boson is the four - component real vector that is transformed under the action of $O(4) \simeq SU(2)_L \otimes SU(2)_R$.
As a result we have the action that consists of three terms $S =  S_K + S_{I} + S_H$, where
\begin{equation}
S_I =  \frac{1}{2}\int d^4x\Bigl(\bar{\Psi}^{{\bf a}}_{i}\gamma^5 \Gamma^5 {\bf H} {\Psi}^{{\bf a}}_{i}  \Bigr)\label{SI0}
\end{equation}
while the pure bare scalar field action is
\begin{equation}
S_H= -\int d^4 x   \frac{M_I^2}{4} {\rm Tr}\, {\bf H}^2 \label{sh_}
\end{equation}


It will be seen below, that the valuable kinetic term for the scalar field arises dynamically through the integration over fermions.
We may rewrite the interaction term as follows:
\begin{eqnarray}
S_I &=& - \frac{1}{2}\int d^4x\Bigl(\bar{L}^{{\bf a}A}_{bi} R^{{\bf a}A}_{ai}H_{ab}+\bar{R}^{{\bf a}A}_{ci}{L}^{{\bf a}A}_{c^{\prime}i}\epsilon_{c^{\prime}b}\epsilon_{ca} H_{ab}  \nonumber\\&& + (h.c.)  \Bigr)
\end{eqnarray}
Here  the scalar field is represented in the form
\begin{eqnarray}
&& H_{ab} = {\bf h}^4 \delta_{ab}+ i \sum_{K = 1,2,3} {\bf h}^K \tau^K_{ab} = H \, U_{\bf h} \label{HSU2}\\&&
 U_{\bf h} = \Bigl[\hat{\bf h}^4 {\bf 1} + {i \sum_{K=1,2,3} \hat{\bf h}^K \tau^K}\Bigr]_{ab}\in SU(2),\nonumber\\&&H = \sqrt{\sum_{k=1,2,3,4}{\bf h}_k^2}, \quad \hat{\bf h}^K = \frac{1}{H}{\bf h}^K\nonumber
\end{eqnarray}
where $\tau^K$ are the Pauli matrices.
Using local $SU(2)_L$ transformation ${L}^{{\bf a}A}_{ai} \rightarrow [U_{\bf h}]_{ab} {L}^{{\bf a}A}_{bi}$ we fix the gauge, in which
\begin{equation}
{\bf H} = H \Gamma^4, \quad H_{ab} = H \delta_{ab}, \quad H = v + h \in {\cal R},\label{unitary}
\end{equation}
where $h$ is the real - valued field of the $125$ GeV Higgs boson while $v$ is the condensate.  Since the obtained expressions are to be considered for the momenta of all particles involved of the order of $M_H$ we may omit the condensate in the interaction term (as it contributes to the zero momentum component $H_{p = 0}$) and arrive at
\begin{equation}
S_I = - \int d^4x\Bigl(\bar{L}^{{\bf a}A}_{1i} R^{{\bf a}A}_{1i}{h}+\bar{R}^{{\bf a}A}_{2i}{L}^{{\bf a}A}_{2i} { h}   + (h.c.)  \Bigr)\nonumber
\end{equation}
 In this form the interaction term coincides with that of \cite{ZetaHiggs} written in Unitary gauge for small distances ($\sim 1/M_H$). For example, quarks of the first generation interact with the real - valued field $h$ as follows:
$S^{(ud)}_I = - \int d^4x\Bigl(\bar{u}^{A}_{L} u^{A}_{R}{h}+\bar{d}^{A}_{R}{d}^{A}_{L} { h}   + (h.c.)  \Bigr)$.

\section{The model at the distances $\gg 1/100$ GeV. The appearance of the top - quark mass.}
\label{Secteff}
We suppose, that the interaction between the composite Higgs boson of the form of Eq. (\ref{SI0}) works for the distances $\sim 1/M_H$, i.e. for the momenta squared of all three participating fields (Higgs field, and the two fermionic fields) $|p^2| \sim M_H^2$. At larger distances $\gg 1/M_H$ the global  $U(12) \otimes O(4)$ symmetry is broken. In addition to the interaction term of the form of Eq. (\ref{SI04}), in particular, the following interaction term is allowed:
\begin{equation}
S^{(4)\prime}_I = \frac{ \alpha^2}{ M_I^2} \int d^4x\Bigl(\bar{L}^{(tb)A}_{aK} t^{A}_{R,K}  \Bigr)\Bigl( \bar{t}^{B}_{R,N} {L}^{(tb)B}_{aN} \Bigr),\label{intftb}
\end{equation}
where $\alpha$ is a dimensionless constant.
$L^{(tb)A}_{aK} = \left(\begin{array}{c}t^A_{L,K}\\b^A_{L,K}\end{array}\right)$ and $t^A_{R,K}$
are the left - handed doublet of the top and bottom quarks and the right - handed singlet of the top quark.
$K = 1,2,3$
is the color index.  Eq. (\ref{intftb}) appears as a result of the integration over the Higgs boson field in the theory with action
$S =  S_K + S_H + S^{\prime}_{I}$, where
\begin{equation}
S^{\prime}_I =  \frac{\alpha}{2}\int d^4x\Bigl(\bar{\Psi}^{{\bf 3}}_{K} \,  \gamma^5 \Gamma^5\, \Big(\hat{\Pi}_+{\bf H} +{\bf H} \, \hat{\Pi}_-\Big) \, {\Psi}^{{\bf 3}}_{K} \Bigr)\label{SI02}
\end{equation}
 Here $\hat{\Pi}_{\pm}$ is the projector that distinguishes the components of $\Psi$ corresponding to the right - handed top quark $t_R$:
\begin{equation}
\hat{\Pi}_{\pm} = \frac{1 \pm \frac{1}{2i} [\Gamma^1,\Gamma^2]\gamma^5 }{2} \times \frac{1-\Gamma^5}{2}\nonumber
\end{equation}

We assume, that at the distances much larger, than $1/M_H$ the interaction term of the form of Eq. (\ref{SI02}) dominates while at the distances of the order of $1/M_H$ the interaction term of Eq. (\ref{SI0}) dominates. We are able to use the action that interpolates between the two:
\begin{widetext}
\begin{eqnarray}
 S_I & = & \frac{1}{2}\int d^4x d^4y d^4z \bar{\Psi}^{{\bf a}}_{i}(x)  \gamma^5 \Gamma^5 {\bf H}(z) {\Psi}^{{\bf a}}_{i}(y) G(x,y,z)\label{SI022} \\ && +  \frac{1}{2}\int d^4x d^4y d^4 z \bar{\Psi}^{{\bf 3}}_{K}(x) \,  \gamma^5 \Gamma^5 \Big(\hat{\Pi}_+{\bf H}(z) +{\bf H}(z) \, \hat{\Pi}_-\Big)  \, {\Psi}^{{\bf 3}}_{K}(y)  \Bigl(\delta(x-z)\delta(y-z)-G(x,y,z)\Bigr),\nonumber \end{eqnarray}
\end{widetext}
where the Form - factor $G$ is introduced. We require, that in momentum space it is given by $G(p,k,q) = \int dx dy dz G(x,y,z) e^{ipx + iky + iqz} = (2\pi)^4\delta(q+p+k) g(p^2,k^2,q^2)$ with the function $g$ that tends to $1$ at $|p^2|\sim |q^2|\sim |k^2|\sim M^2_H$ and to zero if the absolute value of at least one of the three arguments is much smaller, than $[100$ GeV $]^2$. For example, we may choose
$G(p,k,q) = (2\pi)^4\delta(q +p+k) \frac{p^2}{p^2 + M^2}\frac{k^2}{k^2 + M^2}\frac{q^2}{q^2 + M^2}$,
where $M \ll 100$ GeV. In coordinate space this form - factor depends on three scalar parameters $W_1 = (x-z)^2$, $W_2 = (y-z)^2$, $W_3 = (x-y)^2$. Function $G(x,y,z)$ is concentrated within the region $M W_1 \sim M W_2 \sim M W_2 \sim 1$, and decreases fast at $M |W_1|, M |W_2|, M |W_2| \rightarrow \infty$.
It is worth mentioning, that the gauge ($SU(3)_C\otimes SU(2)_L\otimes U(1)_Y$) is to be fixed to define the particular form of the function $G(x,y,z)$.  Notice, that we impose the requirement that the constant $\alpha$ of Eq. (\ref{intftb}) is equal to unity. As a result the interaction between the top - quark and the Higgs boson is given by Eq. (\ref{SI02}) with $\alpha = 1$ at any distances. At the same time the interaction of the other fermions with the Higgs boson is concentrated at the distances $\sim 1/100$ GeV.



In the following we assume that the Unitary gauge Eq. (\ref{unitary}) is fixed that gives $H^{ab} = (v + h)\delta^{ab}$, where $v$ is vacuum average of the scalar field $H$. We denote the Dirac $4$ - component spinors in this gauge corresponding to the SM fermions by $\psi$. We omit angle degrees of freedom to be eaten by the gauge bosons. We take into account only the top quark mass. It follows from Eq. (\ref{intftb}) that
$m_t = \, v$.
The value of $v$ is to be calculated using the gap equation that is  the extremum condition for the effective action as a function of $h$.
The action can be rewritten as
\begin{eqnarray}
S &=&  \int d^4x\Bigl( \bar{\psi}(x)( i \partial \gamma - M)\psi(x) - \bar{\psi}(x) \hat{G}_h \psi(x)\nonumber\\&&- M_I^2 (v+h(x))^2  \Bigr),\label{Smixed}
\end{eqnarray}
where $M$ is the mass matrix. It is diagonal, with the only nonzero component $m_t$.
By $G_h$ the $h$ - depending  operator is denoted. It's action is given by $[\hat{G}_h\xi](x_1) = \int d^4 z d^4 x_2 \,  \xi(x_2)  G(x_1,x_2,z) h(z)$ for all fermions except for the top - quark. For the top quark $\hat{G}_h \equiv h$.


In \cite{ZetaHiggs} the theory with the action of the type of Eq. (\ref{Smixed}) was analysed. The one - loop effective action for the Higgs boson is given by Eq. (3.5) of \cite{ZetaHiggs}. In our notations we have:
\begin{eqnarray}
S[H] &=&  \int {d^4x}  \Bigl( \frac{Z_h^2}{2}\,{\rm Tr}\, H^+(x)(-D^2 w(D^2))H(x) \nonumber\\&&- \frac{Z_h^2}{8}(|H|^2 - v^2)^2 \Bigr),\label{SH}
\end{eqnarray}
where $|H|^2 = \frac{1}{2}{\rm Tr}\, H^+ H$, and $H_{ab}$ is the $2\times 2$ matrix in the representation of Eq. (\ref{HSU2}).
Here the gauge fields of the SM are already restored. As a result the usual derivative $\partial$ of the field $H$ is  substituted by the covariant one $D = \partial + i A$ with the Standard Model $SU(2)_L\otimes U(1)_Y$ gauge field $A = (A^K_{SU(2)}\tau^K, A_{U(1)})$. The action of the gauge group element $e^{i A}$ on the $2\times 2$ matrix $H$ is the  multiplication $H^T \rightarrow \,{\rm exp}\,\Bigl(i A^K_{SU(2)}\tau^K\Bigr) \times H^T \times {\rm exp}\,\Bigl(-i{\rm diag}\,[A_{U(1)},-A_{U(1)}]\Bigr)$. Function $w$  obeys
\begin{eqnarray}
&& w(-p^2) \rightarrow 1 \, (|p^2| \sim [90 \, {\rm GeV}]^2); \nonumber\\&& w(-p^2) \rightarrow N_c/N_{\rm total}=1/8 \, (|p^2| \ll [90 \, {\rm GeV}]^2)\label{wprop}
\end{eqnarray}
Constant $Z_h$ is given by $Z_h^2  =  \frac{N_{\rm total}}{16 \pi^2}{\rm log}\, \frac{\mu^2}{m_t^2}$.

Notice, that effective action of Eq. (\ref{SH}) appears when the given NJL model is considered in zeta regularization or in dimensional regularization, i.e. when {\it all} quadratic divergences are omitted in {\it all} expressions including the one - loop ones. As a result the $1/N$ expansion may be applied. Notice, that in our case the effective action Eq. (\ref{SH}) is valid in the leading order in $1/N_{\rm total} = 1/24$ for the energies $\gg 100$ GeV and in the leading order in $1/N_c = 1/3$ at low energies $\ll 100$ GeV. The expressions for various quantities in zeta - regularization (or dimensional regularization) contain scale parameter $\mu$. We identify this parameter with the typical scale of the interaction that is responsible for the formation of the composite Higgs boson.
Vacuum value $v$ of $H$ satisfies gap equation $\frac{\delta}{\delta h} S[h] = 0$.
It gives \cite{ZetaHiggs}
$M_I^2  = -\frac{N_c}{8 \pi^2 } {m_t^2}\, {\rm log}\frac{\mu^2}{m_t^2} = - \frac{2N_c}{N_{\rm total}} m_t^2 Z_h^2$.
Due to this minus sign bare four - fermion interaction at the high energy scale $\mu \gg m_t$  looks repulsive. However, the appearance of the vacuum average means, that the interaction is actually attractive (see discussion in the Introduction). The minus sign of bare $M_I^2$ is the result of the subtraction of the quadratic divergency in the gap equation made by the accepted zeta (dimensional) regularization.

In order to derive the Higgs boson mass from Eq. (\ref{SH}) we should expand $H_{ab}$ around its vacuum average $H_{ab} = (v + h) \delta_{ab}$  (the angular Goldstone modes and the gauge fields are to be disregarded). Then up to the terms quadratic in $h$ we have the effective lagrangian  $L_h \approx Z_h^2 h \Big((-\partial^2) w(\partial^2) - m_t^2/2\Big)h$. From Eq. (\ref{wprop}) it follows that the propagator $\Big(p^2w(-p^2) - m_t^2/2\Big)^{-1}$ has the only pole at $p^2 = m_t^2/2$. This gives
\begin{eqnarray}
&& M^2_H \approx    m^2_t/2 \approx 125 \, {\rm GeV}\label{MHMT}
\end{eqnarray}


In order to evaluate the gauge boson masses in the given model we should substitute $H$ in the form of Eq. (\ref{unitary}) to Eq. (\ref{SH}) and cannot neglect nontrivial dependence of $w(\|A\|^2)$ on $A$ (we denote $\|A\|^2 =  (A^3_{SU(2)} - A_{U(1)})^2 + [A_{SU(2)}^1]^2 + [A_{SU(2)}^2]^2 $).
 The typical value of $A$ is given by $M_Z$. As a result we may substitute $w(M_Z^2)\approx 1$ instead of $w(\|A\|^2)$. This gives the effective potential for the field $A$.
\begin{eqnarray}
S_A &\approx &\frac{\eta^2}{2}\, \|A\|^2 \approx  \frac{ N_{\rm total}}{16 \pi^2} m_t^2\,  {\rm log}\, \frac{\mu^2}{m_t^2} \, \|A\|^2\nonumber
\end{eqnarray}
That's why we arrive at the following expression for the renormalized vacuum average of the Higgs field $\eta$ (that is to be equal to $\eta = 246$ GeV in order to provide the observed masses of the gauge bosons):
${\eta^2} \approx 2  Z_h^2 v^2 = \frac{2 N_{\rm total}}{16 \pi^2} m_t^2 {\rm log}\, \frac{\mu^2}{m_t^2}$.
From here obtain $\mu \sim 5$ TeV and $M^2_I \approx - [90\,{\rm GeV}]^2$.

\section{Conclusions}
\label{sectconcl}

We suggest, that there is the hidden $U(12)\otimes O(4)$ symmetry behind the formation of the $125$ GeV Higgs boson.   It gives rise to the $U(12)\otimes O(4)$ symmetric four - fermion interaction term at the momenta transfer $\sim 100$ GeV.
The main output of the present paper is the  {\it hypothesis} (that is supported by the considered model) that the Ultraviolet completion of the Standard Model manifests itself already at the distances $\sim 1/M_H$ and the theory may be represented as the theory of only one Majorana spinor $\Psi$ interacting with the Higgs boson field.
The interaction between the fermions and the scalar field at the distances $\sim 1/100$ GeV receives the form of Eq. (\ref{SI0}).

In the opposite limit at the distances $\gg 1/100$ GeV  the  interaction term of Eq. (\ref{SI02}) with $\alpha = 1$ dominates that provides the nonzero mass for the top - quark. The other fermions on this level of understanding are massless. We assume that their masses appear as the perturbations over the suggested pattern.
Altogether this construction provides the effective action, which gives rise to the relation between the Higgs boson mass $M_H$ and the top - quark mass $m_t$ of Eq. (\ref{MHMT}). It is worth mentioning, that the requirement $\alpha = 1$ is essential for this relation to be valid. This prompts, that there is, possibly, more symmetry behind the formation of the $125$ GeV Higgs boson than is noticed in the present paper.

For our approach the use of zeta or dimensional regularization for the derivation of effective action is essential. This is equivalent to the consideration of the given model in usual cutoff regularization with an infinite number of counter - terms added that cancel all quadratic divergences. The use of zeta/dimensional regularization is important for the derivation of  relation $M_H = m_t/\sqrt{2}$. {\it Notice, that in usual cutoff regularization without subtraction of quadratic divergences this relation is not valid even in one - loop.} As it was discussed in the Introduction we accept the {\it hypothesis} that the unknown renormalizable theory to be approximated by the effective NJL model considered in this paper is organized in such a way that due to a certain stability principle all quadratic divergences of the effective model are cancelled within the whole theory. This cancellation may be similar to the cancellation of divergences in vacuum energy of quantum hydrodynamics provided by the whole microscopic theory of a liquid due to the thermodynamical stability of vacuum (for the details see \cite{VolovikKlinkhamer2008,KlinkhamerVolovik2010,Volovik2013}).

It is also worth mentioning, that according to the calculations of Section 3.C of \cite{ZetaHiggs} the particular form of the Form - factor $G$ entering Eq. (\ref{SI022}) may be chosen in such a way, that the Higgs boson branching ratios and its production cross section calculated using the given model match the present experimental data.

The form of the actions of Eqs. (\ref{I4l_}), (\ref{SI0}) prompts the possible origin of forces that provide the compositeness of the $125$ GeV Higgs boson. All participating fermions are arranged in the single spinor. Therefore, we may suppose, that there exist the corresponding gauge field $A$ (its group is the certain extension of the SM gauge group, it may also contain the $SO(3,1)$ spin connection components). Besides, the nontrivial vierbein and its extension may be introduced, i.e. in Eq. (\ref{I4l_}) we may insert
$[E^{\mu}_A]^{{\bf a} j}_{{\bf b} i} \gamma^A$ instead of $\gamma^{\mu}$, where $E_A^{\mu}$ is the $12\times 12$ matrix field that acts on the spinor $\Psi^{\bf b}_j$. This field $E$ is the extention of the notion of vierbein \footnote{Notice, that such an extention of the vierbein appears in certain condensed matter systems (see, for example, \cite{VZ2014_NPB}).}. (The conventional vierbein $f^{\mu}_A$ is the trace mode of $E$: $f^{\mu}_A = \frac{1}{12}[E^{\mu}_A]^{{\bf b} i}_{{\bf b} i}$.) We suppose, that its non - trivial dynamics and/or the non - trivial dynamics of the gauge field $A$ may cause the forces binding the SM fermions in the Higgs boson\footnote{Sure, the dynamics of the trace modes of $E$ should be related to quantum gravity. We may suppose, that the scale of the fluctuations of the trace modes of $E$ is the Plank mass while the scale of the fluctuations of the traceless modes of $E$ is essentially smaller - say, of the order of several TeV.}. We plan to consider this possibility in the other publications.

The author kindly acknowledges numerous discussions with V.A.Miransky and G.E.Volovik.
The work is supported by the Natural Sciences and Engineering Research Council of
Canada.

\end{document}